\documentclass[sigconf, nonacm]{acmart}

\usepackage{caption}
\usepackage{subcaption}

\AtBeginDocument{%
  }

\setcopyright{acmlicensed}
\copyrightyear{2024}
\acmYear{2024}
\setcopyright{acmlicensed}\acmConference[WWW '24 Companion]{Companion Proceedings of the ACM Web Conference 2024}{May 13--17, 2024}{Singapore, Singapore}
\acmBooktitle{Companion Proceedings of the ACM Web Conference 2024 (WWW '24 Companion), May 13--17, 2024, Singapore, Singapore}
\acmDOI{10.1145/3589335.3651961}
\acmISBN{979-8-4007-0172-6/24/05}

\begin{document}

\title{Measuring Arbitrage Losses and Profitability of AMM Liquidity}


\author{Robin Fritsch}
\affiliation{%
 \institution{ETH Zurich}
 \city{Zurich}
 \country{Switzerland}}
\email{rfritsch@ethz.ch}

\author{Andrea Canidio}
\affiliation{%
 \institution{Cow Protocol}
 \city{Lisbon}
 \country{Portugal}}
\email{andrea@cow.fi}



\begin{abstract}
This paper presents the results of a comprehensive empirical study of losses to arbitrageurs (following the formalization of loss-versus-rebalancing by [Milionis et al., 2022]) incurred by liquidity providers on automated market makers (AMMs). We show that those losses exceed the fees earned by liquidity providers across many of the largest AMM liquidity pools (on Uniswap). Remarkably, we also find that the  Uniswap v2 pools are more profitable for passive LPs than their Uniswap v3 counterparts.

We also investigate how arbitrage losses change with  block times. As expected, arbitrage losses decrease when block production is faster. However, the rate of the decline varies significantly across different trading pairs. For instance, when comparing 100ms block times to Ethereum's current 12-second block times, the decrease in losses to arbitrageurs ranges between 20\% to 70\%, depending on the specific trading pair.
\end{abstract}



\maketitle

\section{Introduction}

Automated market makers (AMMs) have emerged as a cornerstone of decentralized finance, holding billions in liquidity and facilitating trillions in total trading volume.
Ever since their inception, the profitability of providing liquidity to AMMs has been a widely discussed research topic -- both in theory and in practice.
The central question remains: Do earnings from trading fees adequately compensate liquidity providers (LPs) on AMMs for the risks they are exposed to?
The answer to this question is crucial, as the long-term sustainability of AMMs, and by extension, much of decentralized finance, relies heavily on LPs receiving sufficient compensation.

While earning trading fees, liquidity providers on AMMs face adverse selection costs. Although any form of market-making activity generally involves such costs, liquidity on AMMs is exposed to a distinct form of losses to arbitrageurs that is unique to this model.
The prevalent type of AMM, the constant function market maker (CFMM), by design, incurs a loss with each price movement on external markets. The intuitive reason is that a CFAMM offers to trade at an outdated (``stale'') price compared to continuous-time off-chain markets at the beginning of each block.
Arbitrageurs capitalize on this price difference, earning profits at the expense of the AMM's LPs.\footnote{In practice, competition between arbitrageurs leads to most of these profits being handed on to validators for the privilege of being the first transaction in a block to interact with the pool.}

In an influential contribution to measuring LP profitability, Milionis et al.~\cite{milionis2022automated} formalized these losses to arbitrageurs by comparing providing liquidity to an AMM with a \emph{rebalancing portfolio} that executes the same trades as an AMM position, but at the external market price. They use a continuous-time model and propose the term \emph{loss-versus-rebalancing} (or short LVR) for the difference between the value of the liquidity position and the rebalancing portfolio.
The concept offers a new angle to measuring LP profitability by removing market risk since the rebalancing portfolio hedges the position's market risk.

Building upon this notion, a liquidity position can be deemed as unprofitable if its earnings from fees fall short of the losses incurred to arbitrageurs.
Note that, throughout this paper, we use the expressions \emph{arbitrage losses}, \emph{losses to arbitrageurs} and \emph{arbitrage profits} interchangeably to refer to the losses that LPs suffer to arbitrage trades due to outdated prices on the AMM, or equivalently, the profits that the arbitrageurs make on these trades. 
In the continuous-time setting without AMM pool fees, this value aligns with LVR.
We extend this concept to the real-world setting characterized by discrete trade times (whenever a new block is appended to the blockchain) and AMM trading fees (similar to the theoretical model in \cite{milionis2023automated}).

In this paper, we empirically study the profitability of liquidity provision using historical data.
In other words, we measure whether earnings from trading fees sufficiently compensate liquidity provided to AMMs for the incurred arbitrage losses.
Our results also quantify the potential benefits of some recently-suggested AMM designs aiming to capture arbitrage profits for liquidity providers (e.g.\ \cite{josojo2022mev}, \cite{mcmenamin2022diamonds} and \cite{canidio2023arbitrageurs}).

We compare the historical earnings from trading fees to the amount of incurred arbitrage losses for the most-traded Uniswap v2 and v3 pools. This covers a considerable fraction of the total liquidity in AMMs, as Uniswap v3 is the market-leading and highest-volume AMM on Ethereum. Uniswap v2, the market leader before the introduction of v3, further maintains a substantial amount of liquidity (about 70\% as much as Uniswap v3, at the time of writing).
Fee earnings are computed based on the historical trading activity within the liquidity pools. To simulate arbitrage losses, we assume that the pools are consistently rebalanced to historical prices on Binance, the most liquid cryptocurrency exchange.

Moreover, this paper also studies a proposed measure to reduce LVR: the reduction of block times.
To do so, we quantify how arbitrage losses depend on the time interval between blocks.

Our analysis reveals several notable findings.
First, we confirm what has previously been mostly known anecdotally:
Fees do not sufficiently compensate for arbitrage losses in most of the largest Uniswap liquidity pools, i.e.\ historically, returns from fees have been smaller than losses to arbitrageur.
This result questions the amount of liquidity (worth billions of USD at the time of writing) currently provided to these AMM pools.
On the other hand, we do find some profitable pools among those with less-traded tokens.

Another noteworthy result is that Uniswap v2 pools, which the more capital-efficient v3 pools have largely superseded, perform remarkably well when comparing fees to arbitrage losses. In particular, the most-traded Uniswap v2 pools are significantly more profitable than the corresponding Uniswap v3 pools for the same pair and trading fee.

Lastly, we observe varying behavior in the relationship between arbitrage losses and blockchain block times across trading pairs.
While faster block times, as expected, reduce losses for all studied pairs, the extent and speed of the decline varies notably.
We find that losses to arbitrageurs are reduced by between 20\% and 70\% with 100ms block times compared to Ethereum's current 12 seconds, depending on the trading pair.

\section{Related Work}

The prevalent type of AMM utilized in DeFi today, i.e. the constant product, or more broadly, constant function market maker, was initially discussed in 2016 \cite{buterin2016dex}, and later formalized in \cite{angeris2021analysis}. Specifically, the AMMs analyzed in this study, namely versions v2 \cite{Adams2020UniswapCore} and v3 \cite{Adams2021UniswapCore} of the Uniswap protocol, are based on these concepts.

The profitability of liquidity on AMMs has been discussed since early in their existence, focusing mostly on so-called ``impermanent loss'' or ``divergence loss'', which compares the development of the value of a liquidity position to holding the liquidity outside the AMM \cite{pintail2019uniswap}.
Empirically, the literature has studied the profitability of liquidity providers (LPs) on automated market makers across various AMMs, token pairs, and time frames.
Heimbach et al.\ \cite{heimbach2021behavior} investigate LP's behavior (and their profitability) on Uniswap v2, while following works, such as \cite{fritsch2021concentrated, heimbach2022risks, loesch2021impermanent}, studied liquidity on Uniswap v3.


In contrast, this paper follows the approach formalized by Milionis et al.\ \cite{milionis2022automated} to measure LP performance versus a ``rebalancing portfolio'' that hedges the price risk of the liquidity position.
The paper also includes an empirical measurement of LVR for the Uniswap v2 WETH-USDC pool.
Cartea et al.\ define \emph{predictable loss} (PL), a metric similar to LVR for quantifying LP losses, and measure it for one Uniswap v3 WETH-USDC pool \cite{cartea2023decentralised}.
This work offers a more comprehensive study, importantly including a wide range of liquidity pools on Uniswap v3 -- which has superseded v2 as the market-leading AMM on Ethereum.
Markout profits of Uniswap v3 liquidity, which are closely related to LVR, have been explored in Twitter and blog posts, see e.g.~\cite{crocswap2022markout}, again for a limited set of pools.
A small subset of the results presented in this paper were previously used in the empirical part of \cite{canidio2023arbitrageurs} which proposes an AMM design utilizing batch auction to prevent LVR.

In follow-up work, Milionis et al.\ extend the model in  Milionis et al.\ \cite{milionis2022automated} to incorporate discrete block times and positive AMM fees \cite{milionis2023automated}.
In particular, their work provides theoretical predictions for the relationship between block times and losses to arbitrageurs, which we measure empirically.
In the results section, we discuss in detail how the theoretical and empirical results compare.

\section{Comparing fees and arbitrage losses}

\subsection{Methodology}

For each liquidity pool we study, we consider a hypothetical liquidity position existing from January 2022 to December 2023. We calculate the fee that such a position would have earned using historical trade data. We also simulate the arbitrage losses that the liquidity position would incur, assuming the AMM pool is consistently rebalanced to Binance prices, the most liquid centralized exchange.

\paragraph{Uniswap v3 liquidity}
In a simple constant-product liquidity pool (such as Uniswap v2 pools), all liquidity positions are available over the entire $[0,\infty]$ price range.
 Uniswap v3 pools, instead, employ the concept of ``concentrated liquidity'', that is, liquidity providers can choose a specific price range they want to provide liquidity to \cite{Adams2021UniswapCore}.
Within each range, liquidity positions act according to a regular constant-product AMM. When a swap occurs, it is executed only against liquidity provided around the price at which the swap occurred. Correspondingly, the swap fees are distributed pro rata among the liquidity available at that price.

For Uniswap v3 pools, we simulate a full-range liquidity position, i.e., a position that provides liquidity the entire price range $[0, \infty]$, as is the case for all Uniswap v2 positions.
Note that, as long as the price stays within the price range of a concentrated liquidity position, it behaves similar to a full-range position.
In particular, the concentrated position earns fees and incurs arbitrage losses in the same way as a full-range position, but requires a smaller amount of reserves to be deposited.
This means that the returns from fees as well as the arbitrage losses relative to the liquidity value scale linearly with the concentration factor of a liquidity position, as long as the position stays in range\footnote{A liquidity position that requires $k$-times less capital will experience $k$-times larges fee returns and arbitrage losses relative to its value.}.

Hence, our results, especially the ratios of fees and arbitrage losses, are also relevant for general liquidity positions in v3 pools as long as they do not go out of range.

\paragraph{Binance prices}

We consider only Uniswap pools whose tokens are traded on Binance.
Moreover, ETH, BTC, USDC, and USDT are traded directly against each other on Binance.
For the remaining pairs, i.e.\ LDO-ETH, LINK-ETH, MATIC-ETH, and UNI-ETH, we derived the Binance prices by combining the prices of the two corresponding USDT pairs.

For all Binance pairs, we retrieve second-by-second price data.
Then, for each Ethereum block, we consider the opening price in the second determined by the block's timestamp.

Finally, note that USDC was not traded on Binance between September 2022 and March 2023.\footnote{During this period, Binance converted USDC and other stablecoins to its own stablecoin, BUSD (see \url{https://www.binance.com/en/support/announcement/binance-to-auto-convert-usdc-usdp-tusd-to-busd-binance-usd-e62f703604a94538a1f1bc803b2d579f}).}
We substitute the missing data by using the corresponding USDT pairs.
We expect the effect of this substitution on the results to be negligible since the prices of USDC and USDT were (almost) equal during the periods we consider and both were available on Binance.

\paragraph{Simulated losses to arbitrageurs}

Using the price data from Binance, for each Uniswap pool, we simulate the amount of arbitrage losses a full-range liquidity position would incur as follows. The liquidity position is deposited into the pool at the beginning of the observation period.
Subsequently, at the time of each block (using the timestamps of Ethereum blocks), we consider the external price on Binance. 
If the difference to the current AMM pool price is sufficient to create a profitable arbitrage opportunity between the pool and the external market (i.e., Binance), we simulate the AMM (and thereby the simulated position) being rebalanced by the optimal arbitrage trade.
That is, arbitrageurs trade with the pool until the marginal price (taking fees into account), equals the Binance price.
In other words, we simulate an AMM that exclusively trades with arbitrageurs.
Note that we do not consider blockchain transaction costs when determining profitable arbitrage opportunities. 
Since we consider that the pool charges a trading fee, the price difference needs to be larger than the fee for a profitable arbitrage opportunity to exist.

The difference between the price at which the arbitrage trade is executed (against the liquidity position) and the external price leads to a profit for the arbitrageurs and a loss to liquidity providers.
For each arbitrage trade, we calculate the size of the loss relative to the value of the liquidity position, and apply the loss to the liquidity position in a compounding manner over time.

\paragraph{Historical fees}

To compute historical returns from fees, we extract all swap transactions from Ethereum events.
For each swap, we retrieve the amount swapped, the fees paid, and the amount of liquidity available at the swap's price (more precisely, at the marginal price after the swap).
Then, by comparing the size of the simulated position to the available liquidity, we can calculate the amount of fees that the simulated position receives.
We put this in relation to the current position value to calculate the relative profit and compound the profit into the liquidity position.\footnote{In practice, fees automatically compound in Uniswap v2 pools while they do not in v3 pools. There, liquidity providers need to repeatedly add earned fees to their position to achieve compounding.} Note that we assume the size of the simulated position to be small compared to the pool size, in the sense that its presence does not affect the behavior of traders and other LPs. 

This method of fee computation makes an implicit assumption, namely that during a swap, the price stays within a price range with constant liquidity.
For Uniswap v3 pools, this can lead to small inaccuracies when swaps cause large price movements, and available liquidity varies across these price ranges.
Besides these potential inaccuracies being non-systematic, such situations are expected to occur only rarely as we consider highly liquid pools on Uniswap where the price impact of single swaps is generally small.\footnote{To illustrate this, we calculate that the difference between assuming liquidity to be constant over a full block instead of over each swap is 0.01\% over 6 months for the WETH-USDC 0.05\% pool. We expect the inaccuracies from assuming the liquidity to be constant during each swap to be even smaller.}

\begin{figure}[ht]
    \centering
    \begin{subfigure}{\columnwidth}
        \includegraphics[width=\columnwidth]{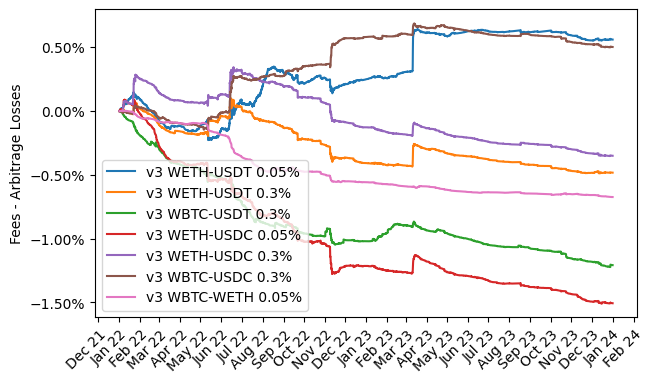}
        \caption{The difference between historical fees and simulated losses to arbitrageurs relative to the liquidity value for a full-range liquidity position. An upwards trend indicates fees being larger than losses.}
        \label{fig:fees_minus_lvr_majors}
    \end{subfigure}%
    
    \begin{subfigure}{\columnwidth}
        \includegraphics[width=\columnwidth]{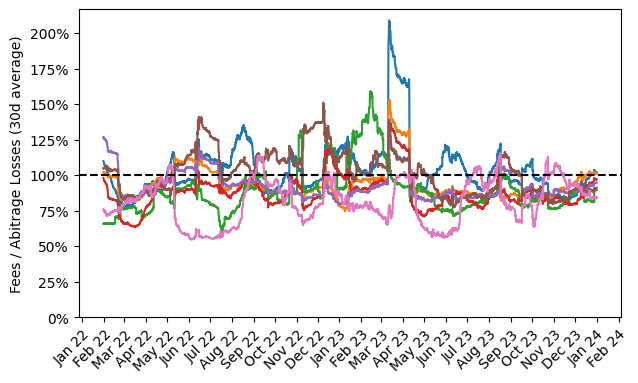}
    \caption{Historical fees relative to simulated losses to arbitrages overtime (30-day moving average).}
        \label{fig:fees_over_lvr_majors}
    \end{subfigure}
    
    \caption{ETH and BTC pairs on Uniswap v3.
    }
    \label{fig:lvr_vs_fees_majors}
\end{figure}

\subsection{Results}

Figure \ref{fig:fees_minus_lvr_majors} shows the cumulative period-by-period difference between fees and arbitrage losses for a simulated full-range liquidity position for the most liquid Uniswap v3 ETH and BTC pools.
Downwards trending lines indicate fees consistently lower than arbitrage losses over the two-year observation period.
We observe this to be the case for most ETH and BTC pools, including the highest-traded Uniswap pool during most of the observation periods, the WETH-USDC 5bp pool.

This fact can also be observed in Figure \ref{fig:fees_over_lvr_majors}, which shows the 30-day moving average of the period-by-period fees earned as a percentage of arbitrage losses for a subset of the pools.
For instance, in the WETH-USDC 5bp pool, the historical returns from fees hover around 80\% of arbitrage losses.

The WETH-USDT 5bp pool (together with the WBTC-USDC 30bp pool) is an exception among the large pools in this regard, roughly breaking even during most times. It also exhibits a brief period in March 2023 when fees reached twice as high as losses due to high trading volumes during the UDSC depeg in March 2023.

\begin{figure}[ht]
    \centering
    \begin{subfigure}{\columnwidth}
        \includegraphics[width=\columnwidth]{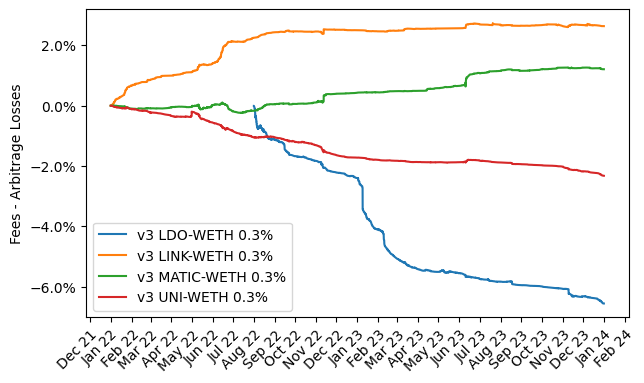}
        \caption{The difference between historical fees and simulated losses to arbitrageurs relative to the liquidity value for a full-range liquidity position. An upwards trend indicates fees being larger than losses.}
        \label{fig:fees_minus_lvr_altcoins}
    \end{subfigure}%
    
    \begin{subfigure}{\columnwidth}
        \includegraphics[width=\columnwidth]{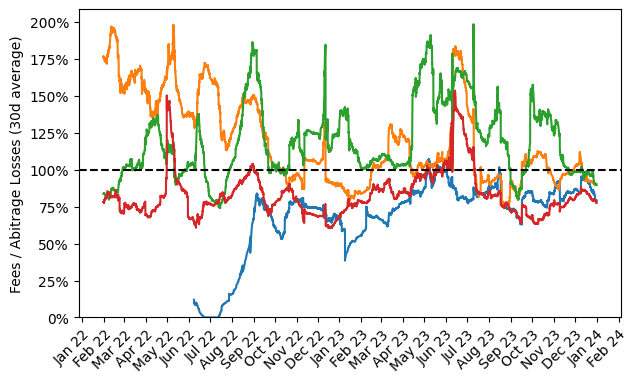}
    \caption{Historical fees relative to simulated losses to arbitrages overtime (30-day moving average).}
        \label{fig:fees_over_lvr_altcoins}
    \end{subfigure}
    
    \caption{Less liquid trading pairs on Uniswap v3}
    \label{fig:lvr_vs_fees_altcoins}
\end{figure}

Figure \ref{fig:lvr_vs_fees_altcoins} shows the corresponding results for several less traded tokens.
For these pairs, fees tend to compensate for arbitrage losses, sometimes even being 50\% higher on average, e.g.\ for the MATIC-ETH and LINK-ETH pools.

\begin{figure}[ht]
    \centering
    \begin{subfigure}{\columnwidth}
        \includegraphics[width=\columnwidth]{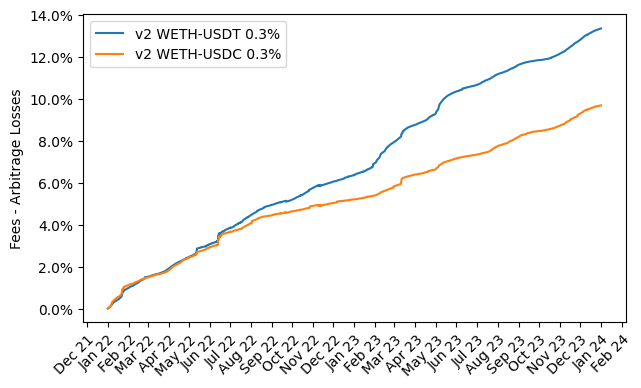}
        \caption{The difference between historical fees and simulated losses to arbitrageurs relative to the liquidity value for a full-range liquidity position. An upwards trend indicates fees being larger than losses.}
        \label{fig:fees_minus_lvr_v2}
    \end{subfigure}%
    
    \begin{subfigure}{\columnwidth}
        \includegraphics[width=\columnwidth]{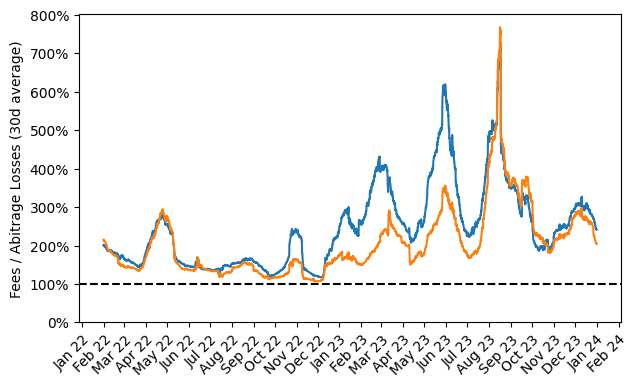}
    \caption{Historical fees relative to simulated losses to arbitrages overtime (30-day moving average).}
        \label{fig:fees_over_lvr_v2}
    \end{subfigure}
    
    \caption{Trading pairs on Uniswap v2}
    \label{fig:lvr_vs_fees_v2}
\end{figure}

Finally, Figure \ref{fig:lvr_vs_fees_v2} shows a significantly different picture for Uniswap v2 pools. Here, fees consistently compensate LPs for arbitrage losses, especially in the second year of the observation period.
Notably, fees are consistently three times larger than losses during this time, as Figure \ref{fig:fees_over_lvr_v2} shows.

\subsection{Discussion}

Our main result is that earnings from fees are smaller than losses to arbitrageurs in the majority of the largest Uniswap pools, currently holding hundreds of millions of USD.
This result raises the question of why LPs nevertheless contribute their capital to these pools. Moreover, the results show the potential impact of mechanisms to reduce LVR in AMMs on the profitability of AMM liquidity.

Finally, the fact that fees relative to the amount of (in-range) liquidity are significantly higher for v2 pools compared to v3 pools, even when comparing pools with equal fee tiers (0.3\%) is arguably surprising.
On both types of AMMs, arbitrage losses are proportional to the liquidity provided. 
The same would be true for retail trading volume if these trades were routed optimally (assuming trading fees are equal).
However, larger pools (e.g., the v3 pools) should see proportionally more volume when fixed costs per swap, such as blockchain transaction fees, are taken into account.
So the ratio of fees to in-range liquidity, and thereby the returns from fees, would be expected to be at least as large in v3 pools as in v2 pools.

One possible explanation for the difference in earnings from fees could lie in an important distinction between Uniswap v3 and v2: Uniswap v3 allows for competition between LPs. Its design enables \emph{active} LPs to move their liquidity around when prices change to optimize their earning from fees, thereby possibly diminishing the share of fees going to \emph{passive} LPs who do not move their liquidity position.
The liquidity position considered in this work is passive, meaning its earnings from fees could potentially suffer from this competition.

\begin{figure}[htp]
    \centering
    \begin{subfigure}{0.9\columnwidth}
        \centering
        \includegraphics[width=0.9\columnwidth]{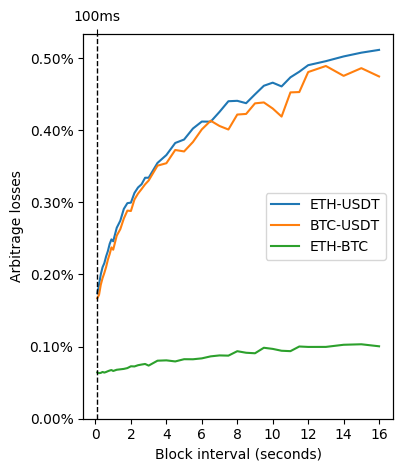}
    \end{subfigure}

    \begin{subfigure}{0.9\columnwidth}
        \centering
        \includegraphics[width=0.9\columnwidth]{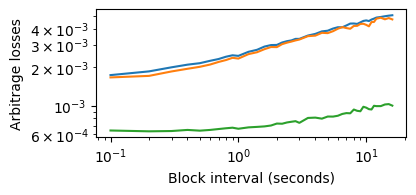}
    \end{subfigure}
    
    \caption{ETH and BTC pairs with a trading fee of 0.05\%: Simulated losses to arbitrageurs relative to the liquidity value for different block intervals (100ms - 16s). The current Ethereum block time is 12s. The lower plots show the same data with logarithmic axes.}
    \label{fig:lvr_by_interval_majors}
\end{figure}

\begin{figure}[htp]
    \centering
    \begin{subfigure}{0.9\columnwidth}
        \centering
        \includegraphics[width=0.9\columnwidth]{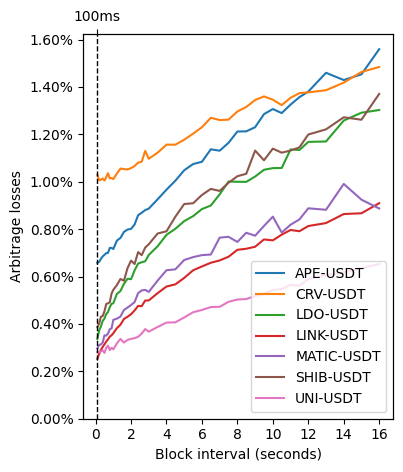}
    \end{subfigure}
    
    \begin{subfigure}{0.9\columnwidth}
        \centering
        \includegraphics[width=0.9\columnwidth]{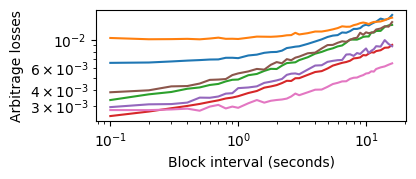}
    \end{subfigure}
    
    \caption{Less-liquid trading pairs with of fee of 0.03\%: Simulated losses to arbitrageurs relative to the liquidity value for different block intervals (100ms - 16s). The current Ethereum block time is 12s. The lower plots show the same data with logarithmic axes.}
    \label{fig:lvr_by_interval_altcoins}
\end{figure}

\section{Losses to arbitrageurs and block times}

The intuition behind shorter block times reducing losses to arbitrageurs is that smaller price differences between blocks lead to smaller arbitrage opportunities. We quantify this effect for block times between 100ms and 16s (and up to 5 min in the appendix).

\subsection{Methodology}
To simulate arbitrage losses under varying block times, we obtain six months of Binance order book data (June 2023 to November 2023)\footnote{Note that Binance offers access to order book data only for perpetual futures, and not for spot trading pairs. Hence, in this section, we consider the perpetual future prices for each of the pairs. See here for a description of Binance's perpetual futures: \url{https://www.binance.com/en/feed/post/168298}}. More specifically, we download all updates to the best bid and ask prices, which allows us to deduce the current bid and ask prices at arbitrary regular time intervals.
Note that we do not consider the amount of liquidity in the order book at these prices. Instead, we implicitly assume arbitrageurs are able to buy and sell sufficient amounts at the best bid and ask prices. 
The rationale is that, with Binance being the most-liquid exchange, its prices are in sync with the overall market, which offers sufficiently deep liquidity.

We then simulate a hypothetical AMM position in a constant-product pool for different block times between 100ms and 15s similarly as described in the precious section:
For each block time, we assume that arbitrageurs rebalance the AMM pool at each every block if it is profitable for them. This is the case if and only if the current price on the AMM taking fees into account is below the current best bid price or above the current best ask price.

We assume the pools to have the following trading fees, which, at the time of writing, represent the most-common fee for these kinds of pairs on the most-used AMMs: 0.05\% for ETH and BTC pairs, and 0.3\% for less-liquid ``altcoin'' pairs.

Note that the results do not depend on the size of the initial liquidity position. Larger positions incur proportionally larger losses to arbitrageurs leading to the same loss relative to the position's value.

\subsection{Results}

Our results (see Figures \ref{fig:lvr_by_interval_majors} and \ref{fig:lvr_by_interval_altcoins}) show varying relationships between arbitrage losses and block times for different trading pairs.
Besides the overall magnitude of the losses, the manner of their decline with shorter block times also varies across pairs.
The reduction in losses to arbitrageurs when block time is 100ms block (compared to Ethereum's 12s), ranges between 20\% and 70\% depending on the pair.
For most token pairs, an acceleration in the reduction trends can be observed for short block times.

The log-log plots in the lower figures show the general dependence on block times. A slope of $c$ in these plots would indicate that arbitrage losses are proportional to the block interval to the power of $c$. The slope is about 1/3 for most pairs and block times above 1s, while it is significantly flatter for blocks times shorter than 1s.

Finally, we see in Figures \ref{fig:lvr_by_interval_majors} and \ref{fig:lvr_by_interval_altcoins} that arbitrage losses begin flattening out for longer block times.
This effect can be seen even more pronounced in Figure \ref{fig:lvr_by_interval_long} in the appendix, which reports results for up to 5 min block times. While the results are somewhat noisy, they generally indicate arbitrage profits leveling off at about twice the current rate at 12s block times for block intervals larger than 2 min.

\subsection{Discussion}

The relationship between arbitrage profits and block times has been studied theoretically in \cite{milionis2023automated}.
For short block times, their theoretical model predicts that arbitrage profits are proportional to the square root of the length of the average block intervals. 
While our empirical findings come close to this model for most pairs and block times larger than 1s, we observe a different regime for block times shorter than 1s.
More precisely, arbitrage profits appear to decline more slowly than the theoretical model would suggest.

The deviating behavior could stem from the differences between the theoretical model and our simulation setup.
Firstly, their model assumes blocks arrive according to a stochastic process while we operate with fixed intervals between blocks. 
Secondly, the asset prices are assumed to follow a geometric Brownian motion whereas base our analysis on historical price data.

On a related note, we remark that when examining arbitrage losses for varying pool fees (simulated in the same way as for block times), the simulation results match the expected inversely proportional relationship predicted by \cite{milionis2023automated} as Figure \ref{fig:lvr_by_fee_all} in the appendix shows.

\section{Conclusion}

To the best of our knowledge, our study provides the most comprehensive measurement of LP losses to arbitrageurs (LVR), spanning multiple token pairs, AMM design, and also block time. We derive several significant and somewhat surprising findings. 
Moreover, it motivates further study in this direction of research.

First, we show that LPs contribute large amounts of liquidity (worth billions USD) to AMMs despite losing to arbitrageurs more than what they make in fees. This striking result begs the question of why LPs contribute liquidity at all. 
Second, the difference in profitability between Uniswap v2 and v3 pools is striking and should be studied further, as it could reveal important insights on how to design AMMs that are profitable (for LPs) and hence sustainable.
Finally, quantifying how LP losses to arbitrageurs change with the intervals between trading opportunities can help to choose block times when designing blockchains.

\begin{acks}
The authors wish to thank Jason Milionis and Ciamac Moallemi for helpful discussions and comments.
\end{acks}

\bibliographystyle{ACM-Reference-Format}
\bibliography{references}


\section{Appendix}

\begin{figure}[H]
    \centering
    \begin{subfigure}{0.9\columnwidth}
        \centering
        \includegraphics[width=0.9\columnwidth]{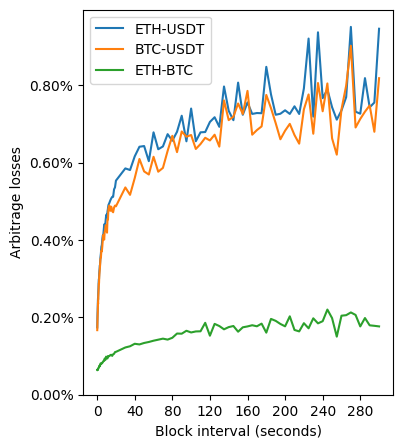}
        \caption{ETH and BTC pairs, trading fee: 0.05\%}
        \label{fig:lvr_by_interval_majors_long}
    \end{subfigure}

    \begin{subfigure}{0.9\columnwidth}
        \centering
        \includegraphics[width=0.9\columnwidth]{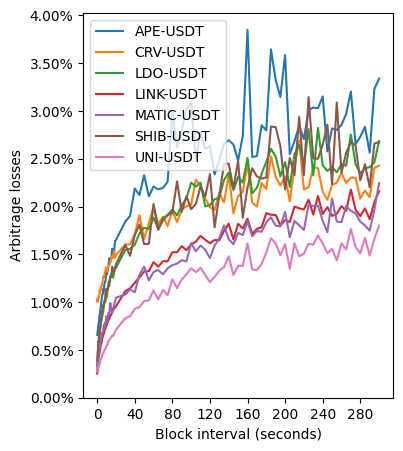}
        \caption{Less liquid pairs, trading fee: 0.3\%}
        \label{fig:lvr_by_interval_altcoins_long}
    \end{subfigure}
    
    \caption{Simulated losses to arbitrageurs relative to the liquidity value for block intervals up to 300s (5 min). The current Ethereum block time is 12s.}
    \label{fig:lvr_by_interval_long}
\end{figure}

\begin{figure}[H]
    \centering
    \begin{subfigure}{0.9\columnwidth}
        \includegraphics[width=0.9\columnwidth]{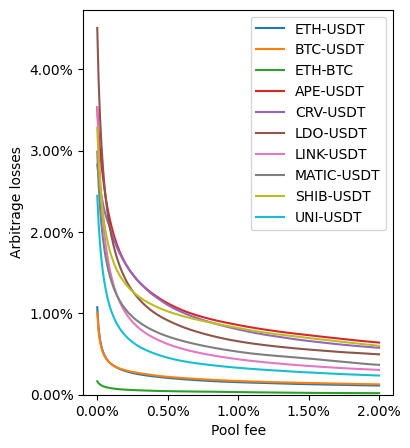}
    \end{subfigure}
    \caption{Simulated losses to arbitrageurs relative to the liquidity value for different pool trading fees.}
    \label{fig:lvr_by_fee_all}
\end{figure}

\end{document}